\documentclass[twocolumn,prb,showpacs,superscriptaddress]{revtex4}
\usepackage{graphicx}%
\usepackage{psfrag}%
\usepackage{dcolumn}
\usepackage{amsmath}
\usepackage{latexsym}

\draft

\begin{document}

\def\bea{\begin{eqnarray}}
\def\eea{\end{eqnarray}}
\def\beq{\begin{equation}}
\def\eeq{\end{equation}}
\def\f{\frac}
\def\k{\kappa}
\def\lh{l_\phi}
\def\sx{\sigma_{xx}}
\def\sy{\sigma_{yy}}
\def\sxy{\sigma_{xy}}
\def\e{\epsilon}
\def\re{\sqrt\epsilon}
\def\ve{\varepsilon}
\def\ex{\epsilon_{xx}}
\def\ey{\epsilon_{yy}}
\def\exy{\epsilon_{xy}}
\def\be{\beta}
\def\D{\Delta}
\def\h{\theta}
\def\t{\tau}
\def\r{\rho}
\def\a{\alpha}
\def\s{\sigma}
\def\kb{k_B}
\def\la{\langle}
\def\ra{\rangle}
\def\nn{\nonumber}
\def\bu{{\bf u}}
\def\bn{\bar{n}}
\def\br{{\bf r}}
\def\up{\uparrow}
\def\dn{\downarrow}
\def\S{\Sigma}
\def\dg{\dagger}
\def\d{\delta}
\def\p{\partial}
\def\l{\lambda}
\def\G{\Gamma}
\def\o{\omega}
\def\g{\gamma}
\def\kv{\bar{k}}
\def\ha{\hat{A}}
\def\hv{\hat{V}}
\def\hg{\hat{g}}
\def\hG{\hat{G}}
\def\hTT{\hat{T}}
\def\noi{\noindent}
\def\a{\alpha}
\def\d{\delta}
\def\p{\partial} 
\def\nn{\nonumber}
\def\r{\rho}
\def\xv{\vec{x}}
\def\rv{\vec{r}}
\def\fv{\vec{f}}
\def\ov{\vec{0}}
\def\vv{\vec{v}}
\def\la{\langle}
\def\ra{\rangle}
\def\e{\epsilon}
\def\o{\omega}
\def\n{\eta}
\def\g{\gamma}
\def\th{\hat{t}}
\def\uh{\hat{u}}
\def\break#1{\pagebreak \vspace*{#1}}
\def\f{\frac}
\def\hf{\frac{1}{2}}
\def\uu{\vec{u}}

\title{A comparative study of two phenomenological models of dephasing in series and parallel resistors} 
\author{Swarnali Bandopadhyay}
\email{swarnali@pks.mpg.de}
\affiliation{
Max Planck Institute for the Physics of Complex Systems, N{\"o}thnitzer Strasse 38, 01187 Dresden, Germany
}
\author{Debasish Chaudhuri}
\email{debc@pks.mpg.de}
\affiliation{
Max Planck Institute for the Physics of Complex Systems, N{\"o}thnitzer Strasse 38, 01187 Dresden, Germany
}
\author{Arun M. Jayannavar}
\email{jayan@iopb.res.in}
\affiliation{
Institute of Physics, Sachivalay Marg, Bhubaneswar 751005, India 
}


\date{\today}

\begin{abstract}
We compare two phenomenological models of dephasing that are in use
 recently. We show that the stochastic absorption model leads to
 reasonable dephasing in series (double barrier) and parallel (ring)
 quantum resistors in presence and absence of magnetic flux. For
 large enough dephasing it leads to Ohm's law.  On the other hand
 a random phase based statistical model that uses averaging over
 Gaussian random-phases, picked up by the propagators, leads to
 several inconsistencies. This can be attributed to the failure
 of this model to dephase interference between complementary
 electron waves each following time-reversed path of the other. 
\end{abstract}

\maketitle

\section {Introduction}
Dephasing is defined as the process by which quantum mechanical
interference is destroyed gradually. 
An electron in a sample may lose its phase memory via interaction
with large number of other degrees of freedom, like a phonon bath or
even due to interaction with all other electrons. While the microscopic
details of such system-bath interactions leading to dephasing is of
interest by itself, we focus on a couple of phenomenological models
that are currently in use to understand how dephasing affects many quantum
interference phenomena in mesoscopic systems. In a double slit setup,
if $\psi_1$ and $\psi_2$ be the two propagating wave functions that
superpose with a phase difference $\phi$, the interference intensity
is $A=|\psi_1|^2+|\psi_2|^2+2Re[\psi_1\psi_2\exp(i\phi)]$. In the 
classical limit of complete decoherence the interference term gets 
totally suppressed and $A=|\psi_1|^2+|\psi_2|^2$.

Most of the mesoscopic samples have dimensions close to the
phase coherence length $l_\phi$, the length scale over which 
the electrons lose their phase memory.  $l_\phi$ can be reduced
by increasing temperature\cite{imry}. An efficient phenomenological model
of dephasing was proposed by B\"uttiker\cite{buttikerVP,buttikerVP2}.
In this model, one attaches `virtual voltage probes' to a system. The probe 
 absorbs phase coherent electrons and in turn reinjects incoherent electrons
back to the system to conserve the overall
unitarity of the system plus the probes. This model uses elastic 
scattering to generate dephasing via
introduction of the virtual voltage probe which carries no net current.
A series of side coupled self-consistent reservoirs, each drawing zero
current, can effectively induce dephasing; moreover this method also
allows one to calculate local chemical potentials and 
temperatures\cite{dib-dabhi}.
The main drawback in B\"uttiker's model is that dephasing occurs 
locally at the point of contact between the system and voltage probes. 
However in a natural sample electrons lose phase memory almost
uniformly due to interaction with other degrees of freedom.

This can be taken into account by adding a spatially uniform imaginary
potential to the Hamiltonian\cite{imagpot} one can introduce uniform
 absorption. The main
problem with this model is that with increase in imaginary potential one
obtains enhanced back reflection, therefore dephasing can not be
 increased monotonically~\cite{kumar,amjIP1,amjIP2}. To avoid
this, an uniform absorption in the coherent wave function can be introduced
via a wave attenuation factor\cite{joshi}. This factor
reduces the wave amplitude by  $\exp(-\a \ell)$ after traversal of a
length $\ell$ in a free propagating region.
Wave attenuation added with a proper incoherent 
reinjection\cite{brouwer} maintaining  the overall unitarity 
can be used to introduce dephasing in the following way.  
The three probe B{\"u}ttiker's model can be mapped into an effective
two terminal geometry by eliminating transmission amplitudes which
explicitly depend on the third (virtual) voltage probe\cite{brouwer}.
In the system, a current $I = I_1 =-I_2$ flows from source
(via lead $1$) to drain (via lead $2$). The chemical potential
of the side coupled voltage probe is adjusted such that no 
current flows through it  i.e. $I_3=0$. 
The unitarity of the $3\times 3 $ $S$-matrix
of the combined system (system plus voltage probe) has been used to eliminate
the transmission coefficients which explicitly depend on the 
virtual voltage probe.
Using Landauer-B\"uttiker
formula~\cite{buttikerVP,buttikerVP2,brouwer,datta} for coherent transport
and eliminating the elements due to the third virtual probe,
the two probe conductance (dimensionless) can be finally expressed 
as\cite{colin,brouwer}
\bea
G &=& \frac{h}{2\,e^2}\,\frac{I}{\mu_1-\mu_2}\nn\\
 &=& T_{21}+\frac{(1-R_{11}-T_{21})\,(1-R_{22}-T_{21})}{1-R_{11}-T_{21}+1-R_{22}-T_{12}}.
\label{reinject}
\eea    
The first term is the transmittance $T_{21}$ from terminal $1$ to terminal 
$2$. The second term is the incoherent reinjection ensuring particle
conservation. 
$R_{11}$ has the meaning of reflectance back
into the first terminal. All other terms in the above expression are
self explanatory.  All these terms are obtainable from a reduced 
$2\times 2 ~ non-unitary S$-matrix which
characterizes an effective two terminal absorbing device.
For further calculation one can use the coefficients of an absorbing
$S$-matrix where absorption takes place uniformly via an wave attenuation
factor as described above.
This model of dephasing is known as
stochastic absorption (SA). For further details one can see
Ref.\cite{colin,brouwer,phiby2}.
 The SA model is shown to produce reasonable
agreement with experimental results\cite{fuhrer}.

In another recently proposed phenomenological model\cite{pala} 
dephasing is introduced via an additive random phase $\phi$ of 
Gaussian distribution to the phase of a propagator after 
it travels a distance $\ell$. In this random phase (RP) statistical model
the following property of Gaussian distributed variables is
utilized. If the mean of the distribution $\la\phi\ra=0$,
$\la \exp(i\phi)\ra=\exp(-\la\phi^2\ra/2)$. Thus if a propagator
picks up an extra Gaussian random phase $\phi$ of mean zero, it loses an
amplitude $\exp(-\la\phi^2\ra/2)$ in an average sense. This will
be equivalent to the loss in amplitude in the SA scheme [$\exp(-\a\ell)$]
if $\la\phi^2\ra=2\a\ell$. Utilizing wave interference it 
was shown\cite{pala} that a Monte-Carlo (MC) averaging over many 
such (perfectly unitary) realizations of random phases is expected to generate 
dephasing. To maintain the Onsager-Casimir reciprocity,
the unitary $S$-matrix of the phase randomizer, is symmetric
and independent of magnetic flux\cite{pala}. Thus a propagator picks up
the same (random) phase in its time-forward and time-reversed path.
It is possible to do an MC averaging of relevant
quantities, like conductance, from individual unitary
processes involving random phase factors to obtain the effect of 
dephasing\cite{pala,zheng}. Other random-phase models, like the 
Lloyd model\cite{dib-kumar1}, are also in use.
Both the above mentioned methods,
stochastic absorption and random phase model, are expected 
to allow one to go from a fully coherent transport to a fully 
incoherent one, continuously, by varying $\a$ from zero to infinity.

In this paper, we use a series and parallel resistor geometry to test
these two models for dephasing. 
We first study a quantum double barrier system (Sec.~\ref{db}). 
The two dephasing models predict completely different results
in the incoherent limit. For the parallel geometry, we choose a quantum
ring in Sec.~\ref{qr}. 
In absence of magnetic flux, we study the impact of dephasing 
on the quantum current magnification (CM)\cite{cm-deo,cm-sw} effect. After
that, in this section, 
we study the two models of dephasing in the context of the 
Aharonov-Bohm (AB) effect. While RP model gives rise to many 
inconsistencies, SA technique produces well behaved 
predictions.  Finally we conclude with some discussions in Sec.~\ref{cnc}.

\begin{figure}[t]
\includegraphics[width=8cm]{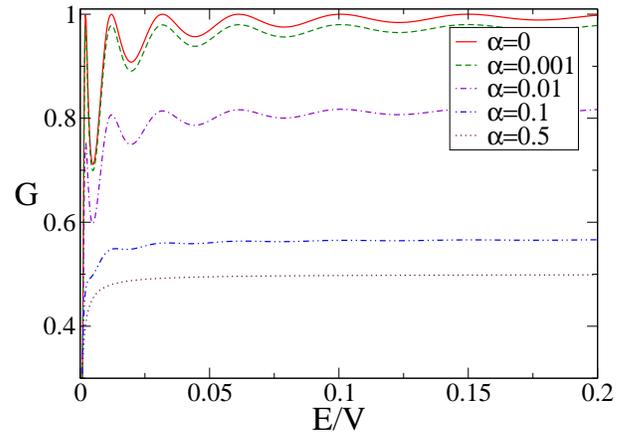}
\caption{Dephasing in quantum double barrier system in 1D obtained from 
stochastic absorption model. 
The data is for two rectangular potential barrier
having same width ($w_1=w_2=0.01$) and strength ($V_1=V_2=V=20)$.
The intermediate distance between two barriers is $s=10$. 
We plot conductance $G$ (in units of $2e^2/h$) as a function of energy $E/V$. 
The SA gives $G=T/2$ in the incoherent limit of large $\a s$, 
$T$ being the transmittance through a single barrier.}
\label{db-sa}
\end{figure}

\section{Double barrier}
\label{db}
Let us assume a double barrier system consisting of two barriers 
characterized by transmittances $T_1$ and $T_2$. 
We consider the expression for two terminal conductance given by $G=T$ 
or equivalently the two terminal resistance ${\cal R}=1/T$.
In the incoherent limit resistances 
should add as in Ohm's law to give the total resistance:  
${\cal R}=1/G=\sum_i 1/T_i$ for a system of $i=1\dots n$ 
barriers connected in series.

To test the SA and RP models of dephasing let us take a one dimensional
double barrier system consisting of two identical rectangular potential 
barriers of strength $V$ and width $w$. Let the two barriers be separated by a 
distance $s$. $k=\sqrt{2mE/\hbar^2}$ is the wave-vector and the
imaginary wave-number is $\k=\sqrt{2m(V-E)/\hbar^2}$.
The $S$-matrix elements for each of the rectangular barriers are,
\bea
t &=& \f{-4ik}{2\k\sinh(\k w)(1-\f{k^2}{\k^2})-4ik\cosh(\k w)},\nn\\
r &=& \f{\k+ik}{\k-ik}\left(-1+t\exp(-\k w)\right)
\eea
and the $S$-matrix is unitary and symmetric. 
In the SA model we assume that the propagator undergoes an attenuation
$\exp(-\a s)$ in a single trip between the barriers.
Thus the various elements of the combined S-matrix required to 
calculate the conductance in SA method are,
\bea
t_{12} = t^2 \exp(-\a s + i k s)/D,\nn\\
r_{11} = r + t^2~ r \exp(-2\a s + 2 i k s)/D
\eea
with $t_{21}=t_{12}$, $r_{22}=r_{11}$. In the above expressions,
$D=1-r^2\exp(-2\a s+ 2 i k s)$.
Using these and Eq.(\ref{reinject}) we find a
loss of oscillation (coherence) in the total conductance
along with a decay in total conductance as a function of 
$\a$ (See Fig.\ref{db-sa}). At large enough absorption $\a s\to\infty$
(practically already for $\a s\ge 5$) the total conductance asymptotically
approaches the value $G\to T/2$, a result expected from the earlier
discussions on incoherent limit ($1/G=1/T+1/T$).  

\begin{figure}[t]
\includegraphics[width=8cm]{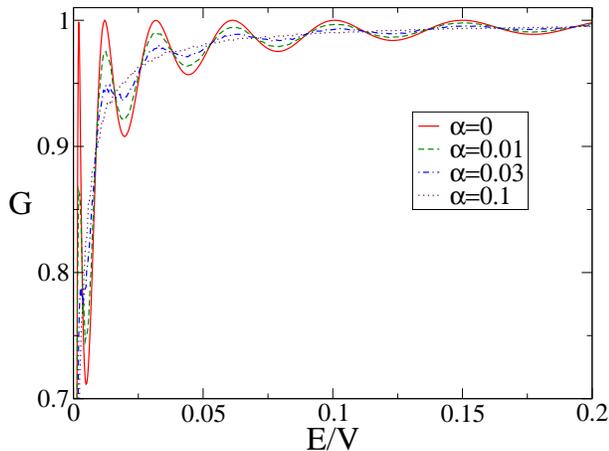}
\caption{Dephasing in quantum double barrier system in 1D using 
random phase model. 
The system parameters are the same as in Fig.\ref{db-sa}.
We plot conductance $G$ (in units of $2e^2/h$) as a function of energy $E/V$. 
The MC simulation predicts $G=T/(2-T)$  in the 
incoherent limit of large $\a s$, $T$ being the transmittance 
through a single barrier.}
\label{db-mc}
\end{figure}

For the implementation of MC averaging over random phases, note that
the transmittance through a single such barrier is
$T=1/[1+(k^2+\k^2)^2~\sinh^2(2\k w)/4k^2\k^2]$ and reflectance is $R=1-T$.
We assume that the propagator picks up
a random phase $\phi$ while traversing the free space ($s$) between the 
barriers, the total transmittance through the double barrier system.
For each realization of the random phase, total transmittance across the
double barrier system is
\beq
T_{12}=\f{T^2}{1-2R\cos(2ks+2\phi)+R^2}.
\eeq
The random phase is assumed to follow a Gaussian distribution of 
mean $\la\phi\ra=0$ and variance $\la\phi^2\ra=2\a s$. 
An average over $500$ realizations
of the random phase is performed at each $\a$ to observe a monotonic
decay of the oscillations (dephasing) in total conductance $G=\la T_{12}\ra$. 
In the limit of large dephasing factor $\a s \to \infty$ 
(practically already for $\a s\ge 1$) $G\to T/(2-T)$ (Fig.\ref{db-mc}). 
This is what one obtains by adding the $S$-matrices of the two
barriers incoherently\cite{datta,zheng}. For two dissimilar barriers
the RP method in the incoherent limit predicts 
$G=T_1 T_2/(1- R_1 R_2)$,  $G= T/(2-T)$ being a special case with
$T_1=T_2=T$.
This result is in clear disagreement with the expectation
of Ohm's law in series resistors in the limit of complete dephasing.

\begin{figure}[t]
\begin{center}
\includegraphics[width=8cm]{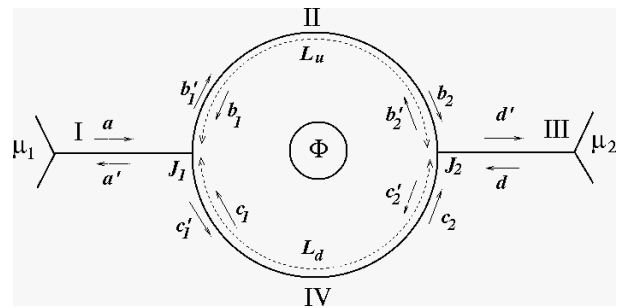}
\end{center}
\caption{
A schematic diagram of the 1D quantum ring considered. $\Phi$ denotes the flux
enclosed. The arrows show counter propagating amplitudes of wave functions.
$\mu_{1,2}$ chemical potentials at two reservoirs connecting the ring at 
the junctions $J_{1,2}$ by two leads.}
\label{ring}
\vskip .5cm
\end{figure}

\section{Quantum ring} 
\label{qr}
We consider a quantum ring (Fig.\ref{ring})
where $L_u, ~L_d$ are the lengths
of the upper and lower arms respectively. The total circumference of the ring
is $L=L_u+L_d$.  
The ring is connected to two leads at junctions $J_1$ and $J_2$.
These connections are described by a scattering matrix $S_c$. For the junction
$J_1$ the outgoing amplitudes $(a',b'_1,c'_1)$ are connected to the incoming
amplitudes $(a,b_1,c_1)$ via $S_c$. Similarly for junction $J_2$, 
$(d' ~b'_2 ~c'_2)^T=S_c(d ~b_2 ~c_2)^T$. The junction S-matrices can be modeled
as\cite{azbel} 
\bea
S_c=\left( \begin{array}{ccc} 
-(p+q) & \re & \re \\
\re & p & q\\
\re & q & p
\end{array}
\right)
\label{sc}
\eea
with $p=[\sqrt{1-2\e}-1]/2$ and $q=[\sqrt{1-2\e}+1]/2$. The $S$-matrix $S_c$
 is characterized by a single parameter $\e$, which can take any value from
$0$ (decoupled) to $0.5$ (strongly coupled). The quantum mechanical continuity
of wave-function and its derivative is attained at the junctions for $\e=4/9$.
Without any lack of generality we use this value for the coupler $\e$ in
our study.

\begin{figure}[t]
\begin{center}
\includegraphics[width=8cm]{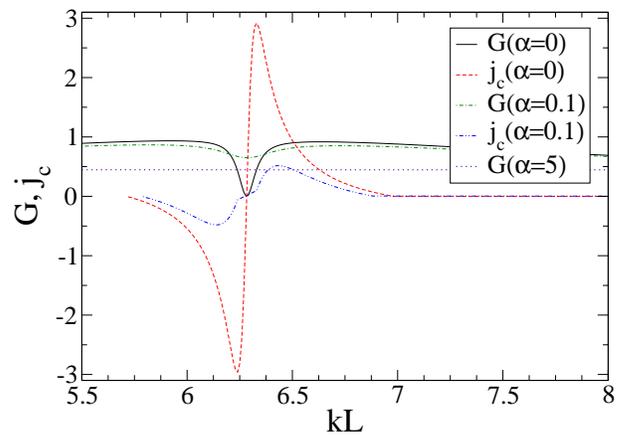}
\end{center}
\caption{Results from stochastic absorption in quantum ring with $L_u/L=0.45$
and $L_d/L=0.55$ in absence of magnetic flux. The coupler $\e=4/9$.
Conductance $G$ (in units of $2e^2/h$) 
and circulating current density $j_c$ (in units of $j_{in}$, see text) 
in a open quantum ring
with changing Fermi momentum $kL$. The amount of circulating current $j_c$ and 
the range of energy over which it is obtained reduces with increase in 
dephasing $\a$. At large enough $\a(=5)$ $G\to\e$.}
\label{sa-cm}
\end{figure}

 The amplitudes $b_1, b_2$ and $b_1', b_2'$ of the upper arm are related by
an $S$-matrix
 \bea
\left( \begin{array}{c}
b_1\\ b_2
\end{array}
\right)
=\left( \begin{array}{cc} 
0 & e^{i(k-\h/L)\,L_u}\,P_u \\
 e^{i(k+\h/L)\,L_u}\,P_u & 0
\end{array}
\right)
\left( \begin{array}{c}
b_1'\\ b_2'
\end{array}
\right).
\label{prop1}
\eea
Similarly for lower arm
\bea
\left( \begin{array}{c}
c_1\\ c_2
\end{array}
\right)
=\left( \begin{array}{cc} 
0 & e^{i(k+\h/L)\,L_d}\,P_d \\
 e^{i(k-\h/L)\,L_d}\,P_d & 0
\label{prop2}
\end{array}
\right)
\left( \begin{array}{c}
c_1'\\ c_2'
\end{array}
\right).
\label{sd}
\eea
Here $k\,L_u$ ($k\,L_d$) denotes the phase picked up by the electron while
traversing the upper (lower) arm of the ring in absence of magnetic flux
 and any dephasing factor. In presence of Aharonov-Bohm flux ($\Phi$) phase 
of the wave-function in upper (lower) arm gets shifted by an amount $\h\,L_u/L$
($\h\,L_d/L$). This phase-shift in two arms add up to give the total flux
piercing the ring {\it i.e.} $\h\,L_u/L+\h\,L_d/L=2\,\pi\,\Phi/\Phi_0$, where
$\Phi_0=h\,c/e$, flux quanta. An extra factor $P_u$ ($P_d$) in the propagator
along the upper (lower) arm introduces dephasing in the system.

\begin{figure}[t]
\begin{center}
\includegraphics[width=8.0cm]{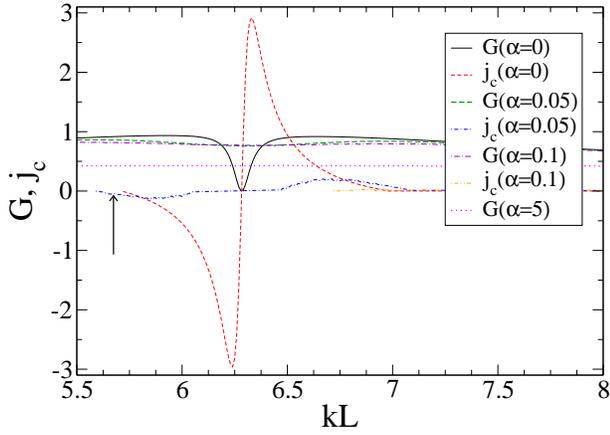}
\end{center}
\caption{Results from random phase model in the quantum ring.
Conductance $G$ (in units of $2e^2/h$)
and circulating current density $j_c$ (in units of $j_{in}$)
in a open quantum ring as a function of 
Fermi momentum $kL$. 
With increase in $\a$ amount of
$j_c$ decreases. Notice the occurrence of $j_c\neq0$ 
for $\a=0.05$ at some Fermi momenta (e.g. as denoted by the arrow)
although for $\a=0$ at the same momenta $j_c=0$. 
At the incoherent limit of large $\a(=5)$ conductance obtained 
from MC averaging is $G<\e$.
The parameters used are $L_u/L=0.45,~L_d/L=0.55$ and $\e=4/9$.
}
\label{mc-cm}
\end{figure}

Before going into the calculation of conductance and impact of dephasing
due to the two dephasing-models we study, let us first discuss what we
expect in the completely incoherent limit. 
The incoming beam transits electrons in the upper (lower) arm with
transmission coefficient $\e$. Hence the upper (lower) arm have 
same resistances at the two junctions $1/\e$. In the incoherent limit  
the resistances in upper (lower) arm  add as in Ohm's law, ${\cal R}_u=2/\e$.
Now these resistances add in parallel leading to the total conductance 
$G=2/{\cal R}_u=\e$.

\begin{figure}[t]
\begin{center}
\includegraphics[width=8.0cm]{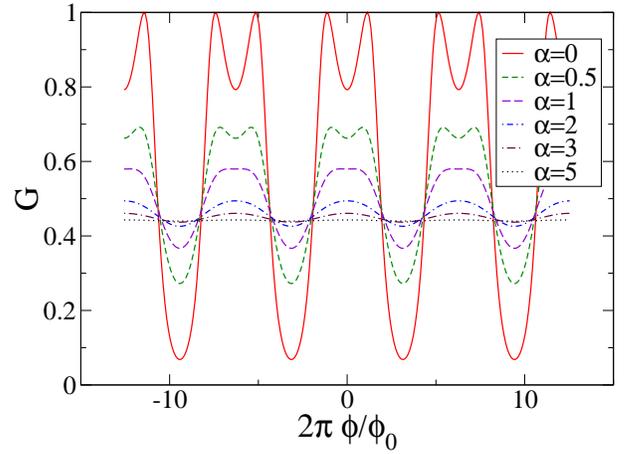}
\end{center}
\caption{Stochastic absorption results in Aharonov-Bohm oscillations.
The ring geometry is $L_u/L=0.45$, $L_d/L=0.55$. The coupler strength $\e=4/9$.
We fixed $kL=5$. The visibility of AB oscillations in $G$ (in units of $2e^2/h$)
clearly dies out with 
increasing $\a$ to asymptotically reach a flux independent conductance
$G=\e$ in the limit of large $\a(=5)$.}
\label{sa-ab}
\end{figure}

In the method of stochastic absorption $P_u=\exp(-\a L_u)$ and
 $P_d=\exp(-\a L_d)$ act as the continuous lossy channels leading to 
dephasing.  Thus combining these $S$-matrices one can obtain
the elements of the effective two terminal `non-unitary' $S$-matrix and  using 
Eq.(\ref{reinject}) find out the conductance $G$ in presence of lossy
channels. In absence of magnetic flux ($\h=0$) we study the conductance
across an asymmetric ring ($L_u\ne L_d$). Throughout this analysis we use
$L_u=0.45$ and $L_d=0.55$ in units of $L=L_u+L_d=1$.
The conductance in such an asymmetric ring is known to
show Fano-resonances as well as Breit-Wigner line shapes as
a function of Fermi energy\cite{cm-deo,cm-pareek,cm-sw}.
The Fano resonances accompany occurrences of circulating current in the ring. 
In presence of transport current through an asymmetric ring system,
depending on Fermi-energy, in one of the arms current can be larger
than the transport current. Thus it is known as current 
magnification\cite{cm-deo,cm-pareek,cm-sw}.
In such a case, to maintain current
conservation, in the other arm, current flows opposite to the bias.
The amount of current flowing opposite to the bias (negative current)
in upper and lower arms of the ring gives the magnitude of circulating
current density $j_c$. We assign positive (negative) sign to $j_c$ for 
clockwise (anti-clockwise) circulating current density i.e. negative 
current density in lower (upper) arm $j_d<0$ ($j_u<0$). 
We measure the current densities in the unit of incident current 
density $j_{in}$. In a small energy interval $dE$ about the Fermi energy, 
the incident current density is $j_{in}= e v \frac{dn}{dE} f(E) dE$, where $f(E)$
is the Fermi distribution function, $\frac{dn}{dE}=2/hv$ is the density of 
states (DOS) in the perfect wire and $v=\hbar k /m$. 
For the zero temperature calculations $f(E)=1$ for occupied states. 
Thus the incident current density becomes $j_{in}=(2e/h) dE$. In
dimensionless units the  current density in upper (lower) arm is
 $j_u/j_{in}=|b'_1|^2-|b_1|^2=|b_2|^2-|b'_2|^2$ 
($j_d/j_{in}=|c'_1|^2-|c_1|^2=|c_2|^2-|c'_2|^2$). 

Occurrence of circulating current
is a purely quantum phase coherence effect. Thus with increasing dephasing
the circulating current density is expected to decay, vanishing completely for 
fully incoherent transport. So we focus on the change in this quantity 
with increasing $\a$ to quantify the degree of dephasing introduced by 
the two dephasing models at hand. 
Fig.\ref{sa-cm} shows the total conductance $G$ and
the circulating current density $j_c$ as a function of Fermi momentum
 ($k L$) in absence
($\a=0$) and presence ($\a=0.1$) of dephasing. It is clear from the figure
that amount of the circulating current density $j_c$ and the range of Fermi
momentum over which it is found, decreases with increasing $\a$. 
It is assumed here that the reinjected current which leads to
classical behaviour does not contribute to circulating current.
 At the $\a\to\infty$ limit it is found that $G\to\e$
[see $G^{sa}(\a=5)$ in Fig.\ref{sa-cm}]. In this limit the circulating
current density is completely absent. Thus predictions from SA is completely 
consistent with our expectations about the incoherent limit.

\begin{figure}[t]
\includegraphics[width=8cm]{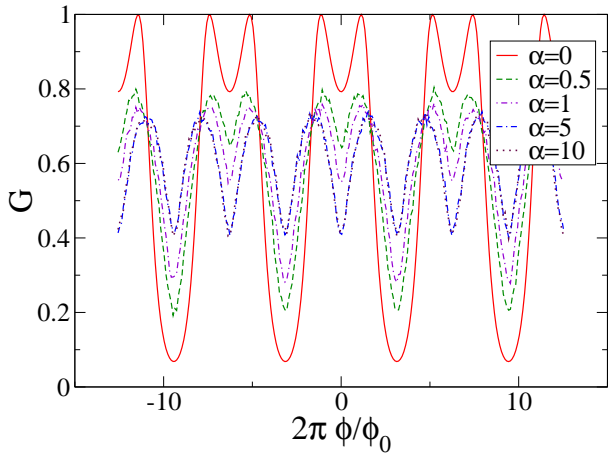}
\caption{Results of random phase model in Aharonov-Bohm 
oscillations. The parameters used are $L_u/L=0.45,~L_d/L=0.55,~kL=5$ 
and $\e=4/9$.
The method clearly fails to reduce the visibility of AB oscillations 
in $G$ (in units of $2e^2/h$) with increasing $\a$ beyond $5$.}
\label{mc-ab}
\end{figure}

In the RP model, on the other hand, a random phase is picked up by the 
propagator while traversing the two arms of the ring. Dephasing in this 
model is introduced by
$P_u=\exp(i\phi_u)$ and $P_d=\exp(i\phi_d)$ such that the phases
obey Gaussian distributions with zero mean and variance
$\s_u^2=2\a L_u$ and $\s_u^2=2\a L_d$. At each realization of the random
phases the $S$-matrices connecting the left and right junctions
 (Eqs.~(\ref{prop1})
and (\ref{prop2})) are unitary and symmetric. Thus they obey Onsager 
reciprocity\cite{datta} at each given realization. In this method one
needs to find the transmittance $T_{12}$ in each realization and perform
an MC averaging over various realizations. 
We averaged over $500$ realizations for each value of $\a$.
From Fig.\ref{mc-cm} it is clear that this method is also capable of reducing
the amount of circulating current density for most of the Fermi momenta. 
However, at some regimes of Fermi momenta, e.g. near $kL=5.6$, we now observe 
non-zero $j_c$ in presence of non-zero $\a$(=0.05) though at the same 
points with $\a=0$ circulating current density was absent (Fig.\ref{mc-cm}). 
This clearly goes against the basic notion of dephasing -- as this is
enhancement of an effect (circulating current) purely attributed to phase
coherence\cite{cm-sw}. SA method is seen to be free from such discrepancies.
However at large enough $\a$(=5) the circulating current density
completely vanishes and the total conductance becomes independent of $k L$.
The 
$\a\to\infty$ limit of $G$ obtained by this method  remains slightly
smaller than that obtained in SA method ($G<\e$, see Fig.\ref{mc-cm}).

Next we switch on the magnetic flux $\Phi$ and study how dephasing
affects the Aharonov-Bohm (AB) oscillations of conductance $G$ as
a function of magnetic flux. The SA absorption method leads to a
monotonic decay in the visibility of oscillations (difference between maxima and
minima) with increasing $\a$ (Fig.\ref{sa-ab}) eventually the conductance
becoming independent of $\Phi$ in the limit $\a L\to\infty$. 
In fact at $\a L=5$ it reaches the incoherent limit of 
$G\to\e$. 
On the other hand, though the RP model leads to
dephasing at small $\a$, it rapidly saturates and fails to kill the
AB oscillation with increasing $\a$ any further.

\begin{figure}[t]
\psfrag{ai}{\Large{$a_i$}}
 \psfrag{a1}{\large{$a_1$}}
 \psfrag{a2}{\large{$a_2$}}
\psfrag{a3}{\large{$a_3$}}
\psfrag{al}{\Large{$\a$}}
\includegraphics[width=8cm]{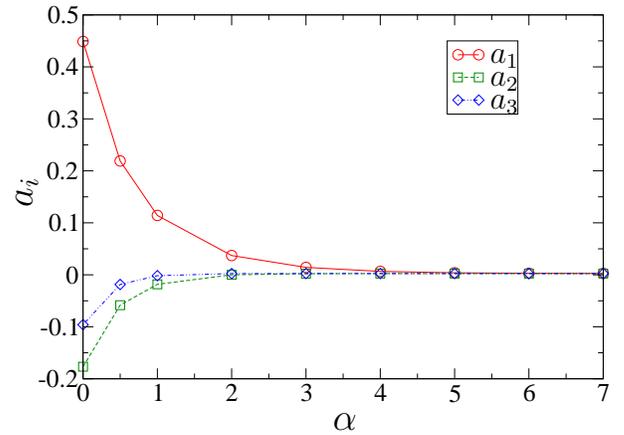}
\caption{First three harmonics $a_i$, $i=1,2,3$ of $G$ as a function of
dephasing factor $\a$. Result from stochastic absorption.}
\label{sa-harms}
\end{figure}

\begin{figure}[t]
\psfrag{ai}{\Large{$a_i$}}
 \psfrag{a1}{\large{$a_1$}}
 \psfrag{a2}{\large{$a_2$}}
\psfrag{a3}{\large{$a_3$}}
\psfrag{al}{\Large{$\a$}}
\includegraphics[width=8cm]{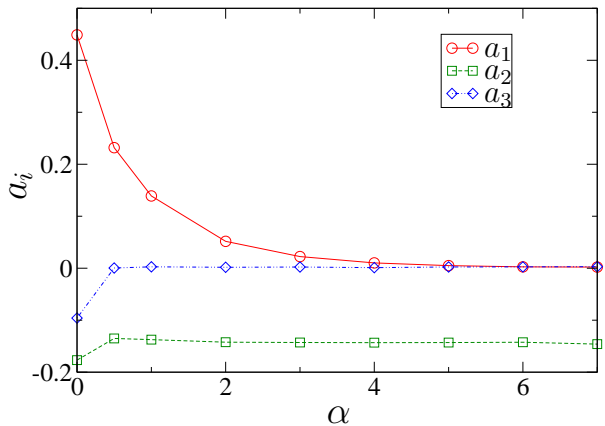}
\caption{First three harmonics $a_i$, $i=1,2,3$ of $G$ as a function of
dephasing factor $\a$. Result from random phase model.}
\label{mc-harms}
\end{figure}

To understand the results, we explicitly extract the various harmonics in the
conductance data presented above. The $n$-th harmonic is
\bea
a_n = \f{1}{\pi}\int_0^{2\pi} G\cos (n\h)d\h.
\eea
The SA method shows fast decay in the various harmonics. In Fig.\ref{sa-harms}
it is clearly seen that the higher is the harmonic the faster it decays to
zero with increasing $\a$. However, the same first three harmonics from MC
data shows a different behavior (Fig.\ref{mc-harms}). 
All the odd harmonics decay to zero, higher harmonics going to zero faster
than the lower one. Whereas the even harmonics show an initial decay followed
by a saturation. It is clear from Fig.\ref{mc-harms} that the second harmonic
$a_2$  saturates to $a_2=-0.14$ with increasing $\a$. Therefore the AB 
oscillation that was predominantly $\Phi_0$ oscillation at $\a L=0$ becomes
predominantly a $\Phi_0/2$ oscillation for $\a L>5$ (Fig.\ref{mc-ab}). In fact, the RP model
fails to kill all the even harmonics in the AB oscillations.

This failure of RP model can be understood in the following way.
Consider two paths which are time reversed version of each other. 
Start from a point to go around the ring (clockwise) once and come back to
origin. One can also travel anticlockwise and come back to the
same point. These are the two time reversed paths.
These two paths together encloses the flux twice. This leads to
the $\Phi_0/2$ oscillation as a function of magnetic flux.
Notice that the phase difference, due to the magnetic flux, between 
these two paths is $4\pi\Phi/\Phi_0$. However, due to the symmetric
nature of the phase randomizing $S$-matrix,  each of these two paths picks 
up the same random phase. Thus in the phase difference, the contribution
from random phase cancels out.
Therefore  phase randomization in the RP model fails to kill even periodicity
($\Phi_0/2n$ with integer n) contributions in one dimensional geometries.
This is true for all even harmonics.

\section{Discussions and conclusion}
\label{cnc}
Studies on series and parallel geometry showed that the SA method leads to
monotonic dephasing (decay in visibility of oscillations or purely coherent
effect like circulating current densities in the ring) with increase in 
dephasing parameter $\a$. 
In the fully incoherent limit of $\a\to\infty$ we recover
Ohm's law, inverse transmittance through individual elements (a barrier
in double barrier case and an arm in quantum ring) behaving as classical
resistances. On the other hand RP model of introducing random
phases to the propagator and averaging over many realizations 
leads to several inconsistencies. For double barrier transmittance,
this leads to an incoherent limit that differs from the classical Ohm's
law. For the transmission through quantum ring, this method generates
circulating current in some regimes of Fermi wave-vectors where there was
no circulating current in absence of phase randomization. 
This behavior is clearly beyond the very principle of dephasing, 
as the circulating current is of purely quantum mechanical origin. 
If the ring is penetrated by magnetic flux,
RP model fails to kill the $\Phi_0/2$ oscillations. This is because
RP model fails to dephase interference between complementary electron
 waves which follow time reversed path of each other.

Thus in conclusion, we find the SA method to be a reliable 
phenomenological technique of introducing dephasing. This systematically
increases dephasing with increasing value of the dephasing parameter $\a$,
finally leading to classical Ohm's law in the limit of large $\a$.
However the RP model fails in many aspects to be a useful phenomenological
model of dephasing.

\section{Acknowledgment}
We thank Martina Hentschel for useful comments on the manuscript.

\end{document}